# Temperature dependence of charge-to-spin conversion in rhombohedral (110) bismuth thin film

Katsuhiro Tatsuoka,[1] Naoki Fukumoto,[1] Shoya Sakamoto,[2] Shinji Miwa,[2] Yuki Fuseya,[3] Junji Fujimoto,[4] Ryo Ohshima,[1,5] Jorge Puebla,[1,5] Yuichiro Ando,[6] and Masashi Shiraishi[1,5 a)]

[1]Department of Electronic Science and Engineering, Kyoto University, Kyoto 615-8510, Japan
[2]The Institute for Solid State Physics, The University of Tokyo, Chiba 227-8581, Japan
[3]Department of Physics, Kobe University, Hyogo 657-8501, Japan
[4]Department of Electrical Engineering, Electronics, and Applied Physics, Saitama University, Saitama 338-8570, Japan
[5]Center for Spintronics Research Network (CSRN), Kyoto University, Kyoto 611-0011, Japan
[6]Department of Physics and Electronics, Osaka Metropolitan University, Osaka 599-8531, Japan

[a)]Author to whom correspondence should be addressed: shiraishi.masashi.4w@kyoto-u.ac.jp

## ABSTRACT

The amplitude of charge-to-spin conversion, namely the spin Hall effect (SHE), in bismuth (Bi) strongly depends on its crystal orientation. The conversion efficiency at room temperature in rhombohedral (110) bismuth is notably large as expected from its large spin-orbit interaction, and such a large SHE is ascribed to the large effective *g*-factor in bismuth [N. Fukumoto *et al*., Proc. Nat. Acad. Sci. **120**, e2215030120 (2023)]. Despite the successful observation of the large conversion efficiency, a more detailed physical mechanism of the SHE in (110) bismuth is still elusive and under debate. In this work, we investigate the temperature dependence of charge-to-spin conversion in an epitaxial Bi(110)/Ni bilayer system using the second harmonic Hall method, revealing that both spin Hall conductivity and spin diffusion length augment with decreasing temperature. This finding suggests that spin scattering in (110) bismuth is dominated by the Elliott-Yafet mechanism, and the charge-to-spin conversion is mainly attributed to skew scattering.

## I. Introduction

Bismuth has attracted significant attention in the field of spin conversion because of its largest spin-orbit interaction among non-radioactive elements [1], and sizable charge-spin interconversion, where spin current is generated orthogonal to the charge current and *vice versa* [2–5]. Whereas sizable spin conversion was observed [6–9], some reports claimed that the conversion efficiency was negligibly small [10–13], setting an ongoing debate in condensed matter physics. Recently, it was demonstrated that the charge-spin interconversion in bismuth strongly depends on its crystal orientation; the interconversion in rhombohedral (110) bismuth [$Bi(110)_R$] is notably large (17–27%), whereas that in $Bi(111)_R$ is almost zero [14,15]. In a previous study, we attributed such anisotropic charge-to-spin conversion to the anisotropic effective *g*-factor in bismuth and possible electron evacuation in $Bi(111)_R$ thin film [15].

Although sizable charge-to-spin conversion was observed in $Bi(110)_R$ at room temperature [15], its detailed underlying physics regarding its spin-dependent scattering mechanisms is still elusive [16]. In the case of platinum (Pt), which is widely and enthusiastically studied in spintronics, crystal quality resulting in a difference in conductivity determines whether the intrinsic or the skew scattering type of extrinsic spin Hall effect (SHE) dominates [17]. The intrinsic SHE in Pt, which reflects the band nature of the material itself, is dominant in a relatively low conductivity regime [17], and it is notable that the intrinsic SHE in Pt can be modulated via electric gating [18]. Meanwhile, the spin Hall conductivity in Pt is enhanced in a high conductivity regime, where the skew scattering is prominent [17]. Thus, investigation of the temperature [19] or crystallinity [20] dependence of the spin relaxation is crucial for studying the scaling relation between the conductivity and the charge-to-spin conversion efficiency for device designing. Considering the chronicle of the SHE mechanisms of Pt, a classification of spin conversion physics, e.g., the SHE and spin relaxation, is significant towards future device applications in other materials exhibiting sizable spin conversion.

In this work, we study the temperature dependence of the charge-to-spin conversion mechanisms in a $Bi(110)_R$/Ni bilayer system using the second harmonic Hall (SHH) method [21–23]. Sizable charge-to-spin conversion of the $Bi(110)_R$ at room temperature is already established in Ref. [15], but the mechanism of the SHE and the spin scattering mechanism of the $Bi(110)_R$ are not discussed. We investigate both mechanisms from the temperature dependence of the spin Hall conductivity and spin diffusion length. It is notable that unwanted thermal effects, which must be considered in the study of charge-to-spin conversion using a material with large thermoelectric response [24], can be quantitatively separated by using the SHH method. It is of great importance compared to the spin torque ferromagnetic resonance (ST-FMR) method [25], which is standard spin-orbit torque (SOT) measurement and is mainly used in Ref. [15]. We evaluate the effective field of damping-like SOT and effective spin Hall conductivity, and interpret further the mechanism of the SHE in $Bi(110)_R$. We corroborate that the skew scattering contribution plays a pivotal role in the SHE in our epitaxially grown $Bi(110)_R$.

## II. Sample Fabrication and Experimental Setup

A Bi/Ni film was deposited on a MgO(001) substrate using molecular beam epitaxy (MBE). The Bi layer was designed to possess a wedged structure and was grown at a temperature below 250 K to obtain well $(110)_R$-oriented structure on the (001)-oriented Ni layer. The film was capped with $SiO_2$(5 nm)/MgO(5 nm) to prevent oxidization. The Bi/Ni film was shaped into a Hall bar structure using Ar ion milling, with a subsequent deposition of Au/Ti electrodes. In the X-ray diffraction (XRD) spectrum of the Bi/Ni film shown in Fig. 1, we observed $(110)_R$ and $(220)_R$ peaks of Bi and a (002) peak of Ni, which supports that the Bi/Ni film used in this study consists of $(110)_R$-oriented Bi and (001)-oriented Ni (see also Ref. [15] for the details). Other negligibly small XRD peaks other than those from the MgO substrate and the sharp streak reflection high energy electron diffraction (RHEED) patterns suggest that intermixing like a formation of $NiBi_3$ interlayer is negligible.

For the investigation of charge-to-spin conversion in $Bi(110)_R$, we follow the second harmonic Hall (SHH) method [21–23] using a Physical Property Measurement System (Quantum Design, PPMS). Figure 2 shows a schematic of the SHH method. We applied an alternating current with an amplitude of $I_0$=5 mA and a frequency of 17 Hz using a current source (Keithley, 6221). The SHH voltage $V_{xy}^{2\omega}$ was measured using a lock-in amplifier (Stanford Research Systems, SR830) while changing the direction $\varphi$ of the in-plane external magnetic field from 0 deg to 360 deg. Measurement temperature ranged from 300 K to 10 K.

## III. Results and Discussion

The $\varphi$ dependence of the SHH signal $R_{xy}^{2\omega} = V_{xy}^{2\omega}/I_0$ is expressed as [22–24,26–28],

$$R_{xy}^{2\omega} = A_{\mathrm{DL}} \cos\varphi + A_{\mathrm{FL}} \cos\varphi \cos 2\varphi + A_{\mathrm{PNE}} \sin 2(\varphi - \delta). \qquad (1)$$

$A_{\mathrm{FL}}$ is the coefficient of the contribution from the field-like spin-orbit torque (SOT) and the Oersted field. The third term $A_{\mathrm{PNE}}\sin2(\varphi-\delta)$ is added to separate the contribution of the planar Nernst effect [26] from unwanted in-plane temperature gradient, where $\delta$ is an angular offset for the in-plane temperature gradient direction. The planar Nernst effect is thermal analogue of the planar Hall effect, and the Hall voltage exhibits $\sin2\theta$ dependence, where $\theta$ is the angle between the magnetization of the ferromagnetic (FM) layer and the in-plane temperature gradient. Since the in-plane temperature gradient, possibly due to the thermal generation at the contact resistance, is the second harmonic component of the applied an alternating current, the Hall voltage due to the planar Nernst effect can be superimposed on the SHH signal $R_{xy}^{2\omega}$. The coefficient $A_{\mathrm{DL}}$ of the contribution from the damping-like SOT is described by taking into account thermal effects due to out-of-plane temperature gradient dependent on the amplitude of the external magnetic field $\mu_0 H_{\mathrm{ext}}$ as,

$$A_{\mathrm{DL}} = -\frac{1}{2}\frac{R_{\mathrm{AHE}}\mu_0 H_{\mathrm{DL}}}{\mu_0(H_{\mathrm{ext}} + H_{\mathrm{K}})} + a_{\nabla T} + b_{\nabla \mathrm{T}}\mu_0 H_{\mathrm{ext}}, \qquad (2)$$

where $\mu_0 H_{DL}$ is the effective field of the damping-like SOT, $R_{AHE}$ is the anomalous Hall resistance, and $\mu_0 H_K$ is the out-of-plane anisotropy field, respectively. $a_{\nabla T}$ and $b_{\nabla T}\mu_0 H_{ext}$ are thermal contributions due to the out-of-plane temperature gradient from the combination of the anomalous Nernst and the spin Seebeck effects, and the ordinary Nernst effect, respectively. Since the sample was grown on an insulating MgO substrate and was set in a vacuum chamber of the apparatus, sizable out-of-plane temperature gradient can arise in the sample. The in-plane external magnetic field is rotated with the angle $\varphi$, hence, the thermal artifacts, such as the anomalous Nernst effect, the spin Seebeck effect, and the ordinary Nernst effect, can be superimposed. Their contribution to the second harmonic signal is considered in the term $A_{DL}\cos\varphi$ in Eq. (1). The anomalous Hall resistance $R_{AHE}$ and the out-of-plane anisotropy field $\mu_0 H_K$ were determined from the first harmonic Hall voltage with sweeping the amplitude of the out-of-plane magnetic field. Figure 3 shows the results of the SHH measurement in Bi(5.5 nm)/Ni(5 nm) at 300 K, 150 K, and 10 K. As shown in Figs. 3(a), (b), and (c), the amplitude of the term $A_{DL}\cos\varphi$ is more than half of the amplitude of the total SHH signal $R_{xy}^{2\omega}$ at all temperatures, which signifies that the term $A_{DL}\cos\varphi$ is dominant in the signal. The contribution from the damping-like SOT was extracted from the $\mu_0 H_{ext}$ dependence of $A_{DL}$ as shown in Figs. 3(d), (e), and (f). The curvature in the $A_{DL}$ as a function of $\mu_0 H_{ext}$ is discernible at all temperatures, which evidences the sizable damping-like SOT, and notably, the concave curves of the $A_{DL}$ indicate positive charge-to-spin conversion efficiencies as previously observed at room temperature [15].

Figure 4(a) summarizes the temperature dependence of $\mu_0 H_{DL}$, extracted from the amplitude of $A_{DL}$ by using Eq. (2), in a series of samples with different $Bi(110)_R$ layer thicknesses. Spin current generated via the spin Hall effect (SHE) in $Bi(110)_R$ is injected into Ni layer and exerts a damping-like SOT on the magnetization. The amplitude of the damping-like SOT is determined by the $\mu_0 H_{DL}$. Thus, the amplitude of the charge-to-spin conversion in $Bi(110)_R$ is expressed as the index $\mu_0 H_{DL}$. As can be seen in Fig. 4(a), the $\mu_0 H_{DL}$ augments with Bi layer thickness, and more importantly, the enhancement is substantial with decreasing temperature especially in thicker Bi systems. For a more precise evaluation, we calculated the effective spin Hall conductivity $\sigma_{SH}^{eff}$ using the following formula [29],

$$\sigma_{SH}^{eff} = \frac{2e\mu_0 M_s t_{FM} H_{DL}}{\hbar E}, \quad (3)$$

where $e$ is the elementary charge, $\mu_0 M_s$ is the saturation magnetization of the sample, $\hbar$ is the reduced Planck constant, $t_{FM}$ is the thickness of the ferromagnetic (FM) layer, and $E$ is the applied electric field that was estimated using the total resistances of the Bi/Ni bilayer measured at zero magnetic field. As shown in Fig. 4(b), $\sigma_{SH}^{eff}$ enhances at low temperature in Bi layer systems thicker than 5.5 nm. In this investigation, we postulate $\mu_0 M_s$=0.61 T [30,31] for Ni at 300 K to be largely unchanged. It is notable that $\mu_0 M_s$ at 0 K is 1.06 times larger than that at 300 K [31], which may give rise to a slight underestimation of $\sigma_{SH}^{eff}$ especially at low temperature, which can results in 0.94(=1/1.06) times smaller $\sigma_{SH}^{eff}$. Meanwhile, the trend of the enhancement of $\sigma_{SH}^{eff}$ at low temperature is salient enough, even though such an underestimation may affect the result. It is also noted that the charge-to-spin conversion efficiency in $Bi(110)_R$ at room temperature was 17–27% [15]. Assuming the conductivity of the Bi layer to be $2.4 \times 10^3\ \Omega^{-1}\mathrm{cm}^{-1}$ (the same value used in Ref. [15]) yields a spin Hall conductivity of $(4.08–6.48)\times10^2\ \Omega^{-1}\mathrm{cm}^{-1}$, and our result at room temperature is consistent with our previous work.

Before considering the mechanism of the spin Hall effect (SHE) in $Bi(110)_R$, we discuss the possible contribution from the orbital Hall effect (OHE) [32,33]. Orbital current, flow of orbital angular momentum, could be generated orthogonal to the charge current via the OHE. More importantly, the orbital current injected into a FM layer exerts a torque on the magnetization. Hence, a contribution from the OHE could be superimposed on $\mu_0 H_{DL}$ or $\sigma_{SH}^{eff}$. Here, it is notable that the torque from the OHE strongly depends on the orbital-to-spin conversion efficiency in the FM material. Ni is known for its orbital-to-spin conversion, and in fact, the conversion efficiency is much larger than that of Fe or $Ni_{80}Fe_{20}$ (Py) [34,33,35]. Thus, a sizable torque from the OHE in $Bi(110)_R$/Ni system may be superposed. Nevertheless, the contribution of the OHE from $Bi(110)_R$ can be negated because comparable charge-to-spin conversion efficiencies were experimentally observed in $Bi(110)_R$/Ni (17%) and $Bi(110)_R$/Fe (34%) using the spin torque ferromagnetic resonance (ST-FMR) measurement [15]. Indeed, if a sizable OHE in $Bi(110)_R$ takes place, the charge-to-spin conversion in $Bi(110)_R$/Ni would be greater than that in $Bi(110)_R$/Fe. However, this is not the case, which buttresses our assertion that the OHE in Bi does not significantly manifest itself. Furthermore, the orbital Hall conductivity in Bi is theoretically calculated using the Kubo formula and is generally expected to be three times smaller than the spin Hall conductivity in Bi [36]. This calculation also supports the negligibly small contribution from the OHE in $Bi(110)_R$/Ni. Hence, the observed $\mu_0 H_{DL}$ or $\sigma_{SH}^{eff}$ is dominantly attributed to the charge-to-spin conversion in $Bi(110)_R$/Ni, not the charge-to-orbital conversion.

Since the Gilbert damping parameter $\alpha$ at room temperature depends on Bi layer thickness [15] and the observed $\mu_0 H_{DL}$ or $\sigma_{SH}^{eff}$ depends on the Bi layer thickness as shown in Fig. 4, the observed charge-to-spin conversion is not ascribable to interfacial effects, such as the Rashba-Edelstein effect. Furthermore, as demonstrated in Ref. [15], the observed charge-to-spin conversion is dominated by the SHE in $Bi(110)_R$, not the self-induced SOT in Ni layer. Assuming ideal spin current transparency at the Bi/Ni interface and thickness-independency of the Bi layer resistivity $\rho_{Bi}$, the Bi thickness $t_{Bi}$ dependence of $\sigma_{SH}^{eff}$ is derived as [29,37,38]

$$\sigma_{SH}^{eff} = \sigma_{SH}\left[1 - \mathrm{sech}\left(\frac{t_{Bi}}{\lambda_{Bi}}\right)\right], \quad (4)$$

where $\sigma_{SH}$ is the spin Hall conductivity and $\lambda_{Bi}$ is the spin diffusion length in the Bi. Figure 5(a) shows the Bi thickness dependence of $\sigma_{SH}^{eff}$ and the fitting results at each temperature using Eq. (4). The estimated $\sigma_{SH}$ and $\lambda_{Bi}$ for various temperatures from 10 to 300 K are shown in Figs. 5(b) and (c), respectively. As mentioned above, we assume the Bi layer resistivity $\rho_{\mathrm{Bi}}$ is thickness-independent. Albeit it is significant to characterize thickness dependence of the resistivity of $\mathrm{Bi}(110)_{\mathrm{R}}$ single layer, we note that crystal growth of $\mathrm{Bi}(110)_{\mathrm{R}}$ on insulating substrates was difficult at this stage, and in fact, merely $\mathrm{Bi}(111)_{\mathrm{R}}$ can be grown on them. Hence, the analysis of the thickness dependence of $\sigma_{\mathrm{SH}}^{\mathrm{eff}}$ is carried out under the assumption that $\rho_{\mathrm{Bi}}$ is identical within the thickness. Regarding the interfacial spin transparency, the spin memory loss at $\mathrm{Bi}(110)_{\mathrm{R}}$/Ni interface may not be negligible because materials with large spin-orbit interaction are used. The quantitative evaluation of the spin memory loss in the system is significant as well. However, the spin memory loss is most likely independent of Bi layer thickness $t_{\mathrm{Bi}}$. In theory, whereas the absolute values of $\mu_0 H_{\mathrm{DL}}$, $\sigma_{\mathrm{SH}}^{\mathrm{eff}}$ and $\sigma_{\mathrm{SH}}$ are affected by the spin memory loss, the value of estimated $\lambda_{\mathrm{Bi}}$ is not affected. Thus, the following discussion about the mechanism of the charge-to-spin conversion in $\mathrm{Bi}(110)_{\mathrm{R}}$ with assuming that the spin transparency is ideal allows precise estimation of $\lambda_{\mathrm{Bi}}$. The $\sigma_{\mathrm{SH}}^{\mathrm{eff}}$ noticeably increases at low temperature, which is unlike polycrystalline Bi-Sb alloys, where the skew scattering contribution is weak [39], and $\lambda_{\mathrm{Bi}}$ gradually increases with decreasing temperature. The known dominant spin relaxation mechanisms in condensed matter are the Elliott-Yafet type and the D'yakonov-Perel' type [20]. In the Elliott-Yafet mechanism, the spin lifetime $\tau_s$ is proportional to the momentum relaxation time $\tau_p$, which results in the spin diffusion length to be proportional to $\tau_p$. In contrast, in the D'yakonov-Perel' mechanism, $\tau_s$ is proportional to $1/\tau_p$ and the spin diffusion length is independent of momentum relaxation time $\tau_p$. Note that mobilities of both electrons and holes in bismuth are strongly enhanced with decreasing temperature [40,41] and the augmentation of the mobility is more significant than the decrease of effective mass with decreasing temperature [42]. Since the mobility is proportional to $\tau_p$ and the spin diffusion length $\lambda_{\mathrm{Bi}}$ increases with decreasing temperature as shown in Fig. 5(c), the spin relaxation in our $\mathrm{Bi}(110)_{\mathrm{R}}$ is likely governed by the Elliott-Yafet mechanism, which is typical for a single metal with spatial inversion symmetry.

It was reported that the extrinsic contribution can exceed the intrinsic contribution of spin Hall effect (SHE) in bismuth [16]. If the spin relaxation in $\mathrm{Bi}(110)_{\mathrm{R}}$ is dominated by the Elliott-Yafet mechanism and the momentum relaxation time $\tau_p$ increases with decreasing temperature, the charge-to-spin conversion can be ascribed to skew scattering, whose contribution to the spin Hall conductivity $\sigma_{\mathrm{SH}}$ is proportional to $\tau_p$ [16]. The relation between the spin Hall conductivity $\sigma_{\mathrm{SH}}$ and the spin diffusion length $\lambda_{\mathrm{Bi}}$ is plotted in Fig. 5(d), where $\lambda_{\mathrm{Bi}}$ is proportional to $\tau_p$. This result is consistent with the dominance of skew scattering in our $\mathrm{Bi}(110)_{\mathrm{R}}$. Simultaneously, this result substantiates that the side-jump contribution is not dominant [43]. The sizable skew scattering contribution is manifested in Bi due to its large mobilities and large spin-orbit interaction.

While the linear relation shown in Fig. 5(d) implies the dominant skew scattering contribution in Bi, the intrinsic contribution can be superposed to a certain extent on the charge-to-spin conversion in Bi. The band structure in bulk bismuth at the $L$-point is determined by the Wolff Hamiltonian, which is essentially equivalent to the Dirac Hamiltonian [1,44]. The intrinsic SHE simply originates from the interband effect of the Dirac electrons at the $L$-point [1,44] and is irrelevant to thermal excitation. Thus, the intrinsic SHE for the Dirac electrons is finite in the low temperature limit. The band gap for electrons at the $L$-point is ~14-35 meV [42], whereas that for holes at the $T$-point is at least ~200 meV [45]. Moreover, carriers in $Bi(110)_R$ thin films are dominated by electrons [46], and electron transfer from the Ni underlayer into the Bi layer may occur, since the electron carrier density is significantly different between ferromagnetic metal Ni and semimetal Bi. Hence, the band gap shrinking at the $L$-point occurs with decreasing temperature due to the band inversion [42], which yields the enhancement of the intrinsic SHE with decreasing temperature. Although the intrinsic SHE can be superimposed on the enhancement of the spin Hall conductivity $\sigma_{SH}$ with decreasing temperature, the results shown in Fig. 5(d) strongly suggests that the skew scattering governs the charge-to-spin conversion in our $Bi(110)_R$ film. Meanwhile, future study decoupling possible superposition of the intrinsic contribution is significant considering temperature-dependent band shrinking for electrons in Bi at low temperature [42]. We note that a supporting result to our assertion is that in Bi-Sb alloys, where the skew scattering contribution is negligible and the intrinsic SHE dominates the charge-to-spin conversion, the spin Hall conductivity $\sigma_{SH}$ does not increase with decreasing temperature [39]. Hence, we deduce that the observed charge-to-spin conversion in $Bi(110)_R$ is mainly ascribed to the skew scattering.

## IV. Conclusion

In conclusion, we measured the temperature dependence of the effective field of the damping-like spin-orbit torque (SOT) $\mu_0 H_{DL}$ and the effective spin Hall conductivity $\sigma_{SH}^{eff}$ in epitaxial $Bi(110)_R$/Ni using the second harmonic Hall (SHH) method. Both $\mu_0 H_{DL}$ and $\sigma_{SH}^{eff}$ increased with decreasing temperature. The increase of the spin diffusion length $\lambda_{Bi}$ with decreasing temperature suggests that spin relaxation in $Bi(110)_R$ is dominated by the Elliott-Yafet mechanism. The enhancement of the spin Hall conductivity $\sigma_{SH}$ at low temperature and the linear relation between $\sigma_{SH}$ and $\lambda_{Bi}$ implies that the SHE in $Bi(110)_R$ is dominated by the skew scattering, and the high-purity sample negates the side-jump mechanism. Meanwhile, the intrinsic contribution due to the band gap shrinking at the $L$-point is expected to be small compared to the skew scattering contribution. The mechanism of the charge-to-spin conversion in $Bi(110)_R$, which was observed to be sizable at room temperature using the spin torque ferromagnetic (ST-FMR) measurement [15], was shed light on from its temperature dependence.

## ACKNOWLEDGEMENTS

We acknowledge E. Wei and W. Gao for experimental assistance. This work was partially supported by Japan Society for the Promotion of Science (JSPS) Grant-in-Aid for Scientific Research (A) (23H00268), MEXT Initiative to Establish Next-generation Novel Integrated Circuits Centers (X-NICS) Grant Number JPJ011438, and the Spintronics Research Network of Japan (Spin-RNJ).

## DATA AVAILABILITY

The data that support the findings of this article are openly available at Ref. [47]. The raw dataset for the second harmonic Hall (SHH) method is stored, that are the first and SHH traces with in-plane external magnetic field and the anomalous Hall traces with out-of-plane external magnetic field.

**Figure captions**

FIG. 1. X-ray diffraction pattern (Cu-$K_\alpha$ radiation) of the Bi/Ni film. Peaks at 38.5° and 43.1° were obtained from the MgO(001) substrate in $K_\beta$ radiation and $K_\alpha$ radiation, respectively. Taken from Ref. [15].

FIG. 2. Schematic of the second harmonic Hall (SHH) method. An alternating current $I^\omega$ is applied to the $Bi(110)_R$/Ni Hall bar and the SHH voltage $V_{xy}^{2\omega}$ is measured. Symbols + and − indicate the configuration of the measurement directions for $I^\omega$ and $V_{xy}^{2\omega}$. The in-plane external magnetic field $\mu_0 H_{ext}$ is applied, where $\varphi$ is the angle between $I^\omega$ and $\mu_0 H_{ext}$.

FIG. 3. Results of the second harmonic Hall (SHH) measurement in Bi(5.5 nm)/Ni(5 nm). The in-plane angular dependence of the SHH signal $R_{xy}^{2\omega}$ with an external magnetic field of $\mu_0 H_{ext}$=1 T measured at (a) 300 K, (b) 150 K, and (c) 10 K, respectively. Offset resistances were removed. The white dots represent the measurement results, and the black lines represent the fitting curves using Eq. (1), respectively. The red solid lines denoted as DL, the blue dashed lines denoted as FL, and the green dotted lines denoted as PNE represent the first, second, and third terms in Eq. (1), respectively. The magnetic field dependence of the coefficient $A_{DL}$ in Eq. (1) derived from the fitting of the SHH signals at (d) 300 K, (e) 150 K, and (f) 10 K, respectively. The white squares with standard deviation are the fitting results of the SHH signals and the black lines are fitting curves using Eq. (2). The results at low external magnetic field, shown with transparent squares, are excluded from the fitting using Eq. (2). The red solid lines denoted as DL SOT (that means damping-like spin-orbit torque), the blue dashed lines denoted as ANE+SSE (that means anomalous Nernst effect and spin Seebeck effect), and the dotted green lines denoted as ONE (that means ordinary Nernst effect) represent the first, second, and third terms in Eq. (2), respectively.

FIG. 4. Temperature dependence of (a) the effective field of damping-like spin-orbit torque $\mu_0 H_{DL}$ and (b) the effective spin Hall conductivity $\sigma_{SH}^{eff}$, obtained in the Bi($t_{Bi}$)/Ni(5 nm) systems, respectively. Numbers in parentheses in the legend are the nominal thicknesses of the Bi layer $t_{Bi}$ in nanometers. Data for the Ni(5 nm) single layer, namely Bi(0) with right-facing black triangles, at 50 K and 10 K are not shown since the fittings of the harmonic signals were not successfully performed. Data for Bi(9.5) (circle with a brown frame) at 10 K is also not shown with the same reason. The measurements at 30 K are conducted for Bi(9.5) (circle with a brown frame), Bi(7.5) (upward-facing purple triangle), and Bi(5.5) (downward-facing triangle with a green frame).

FIG. 5. (a) Plot of the effective spin Hall conductivity $\sigma_{SH}^{eff}$ against the nominal Bi layer thickness at various temperatures and the corresponding fitting curves using Eq. (4). Temperature dependence of (b) the spin Hall conductivity $\sigma_{SH}$ and (c) the spin diffusion length of the Bi layer $\lambda_{Bi}$ derived from the fitting results in (a). (d) The relationship between $\sigma_{SH}$ and $\lambda_{Bi}$. The red dashed line is a fitting line without considering errors.

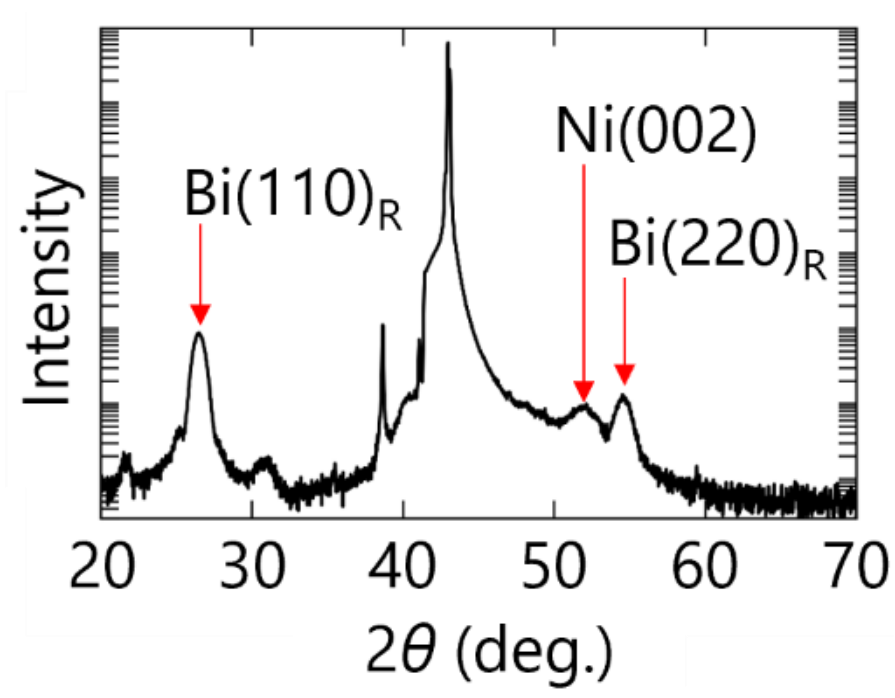

Fig. 1. Tatsuoka *et al*.

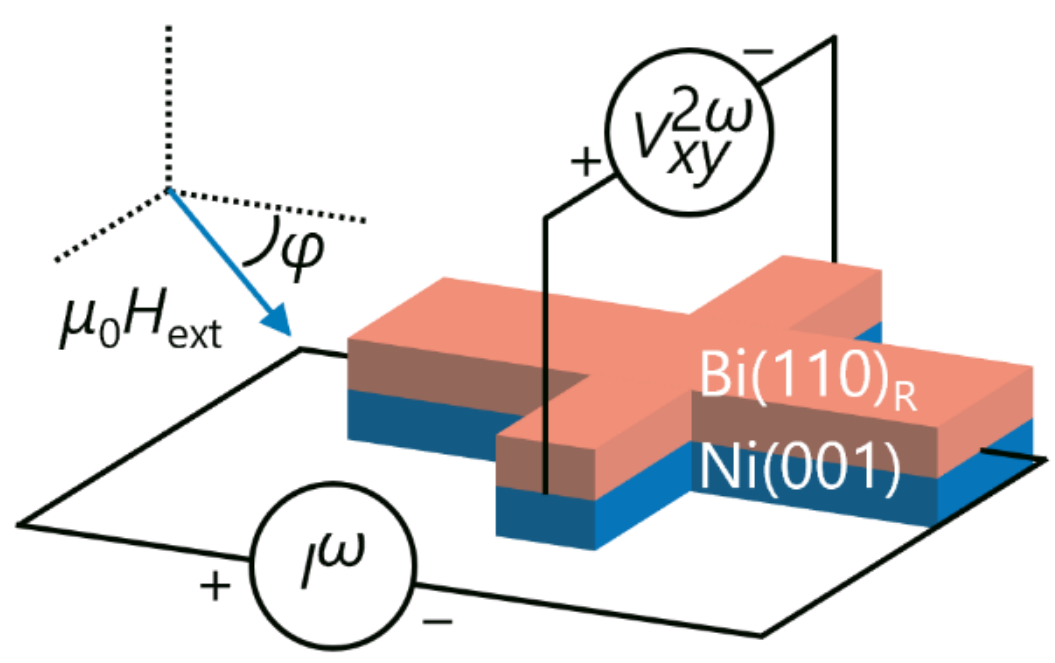

Fig. 2. Tatsuoka *et al*.

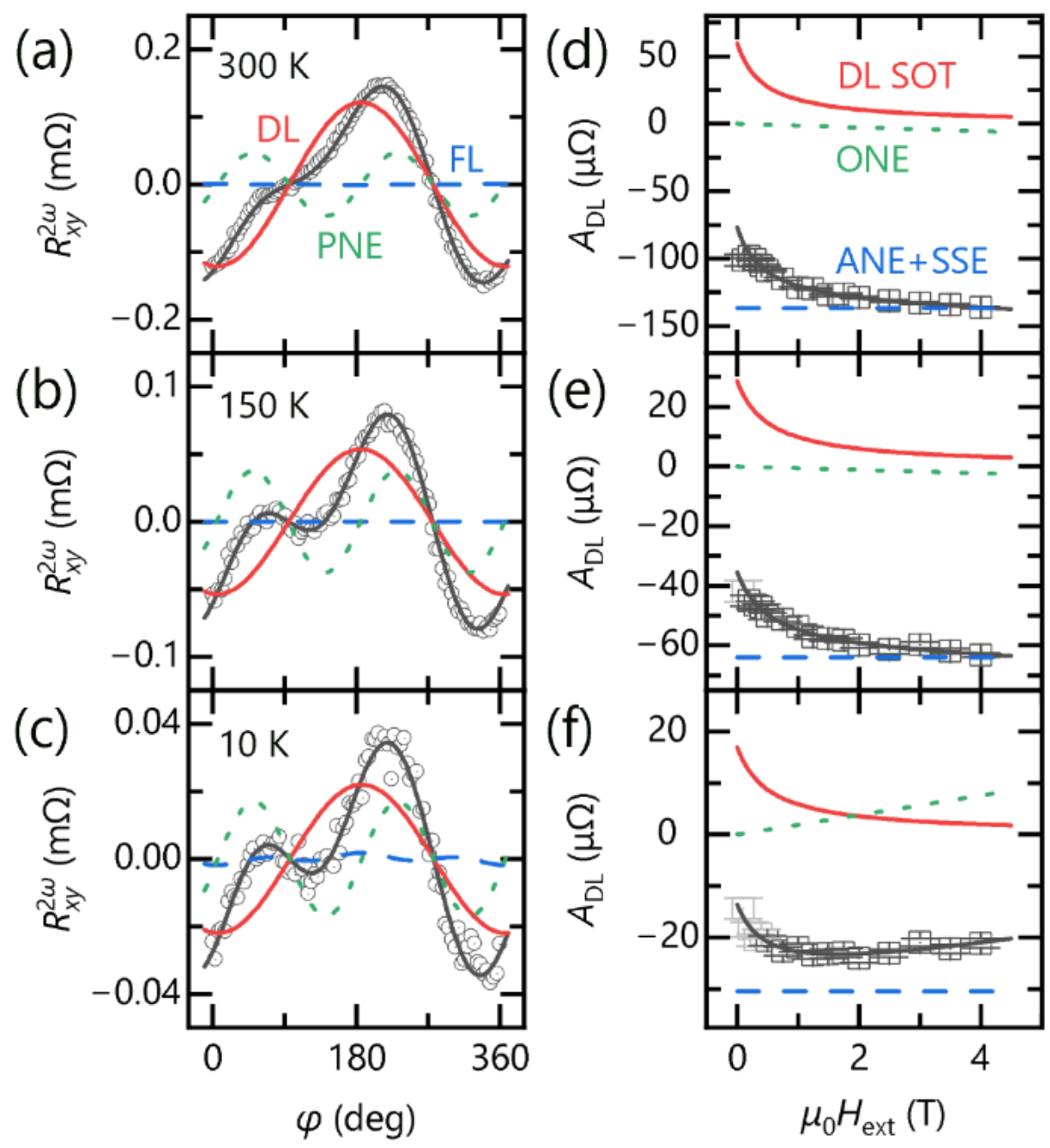

Fig. 3. Tatsuoka *et al*.

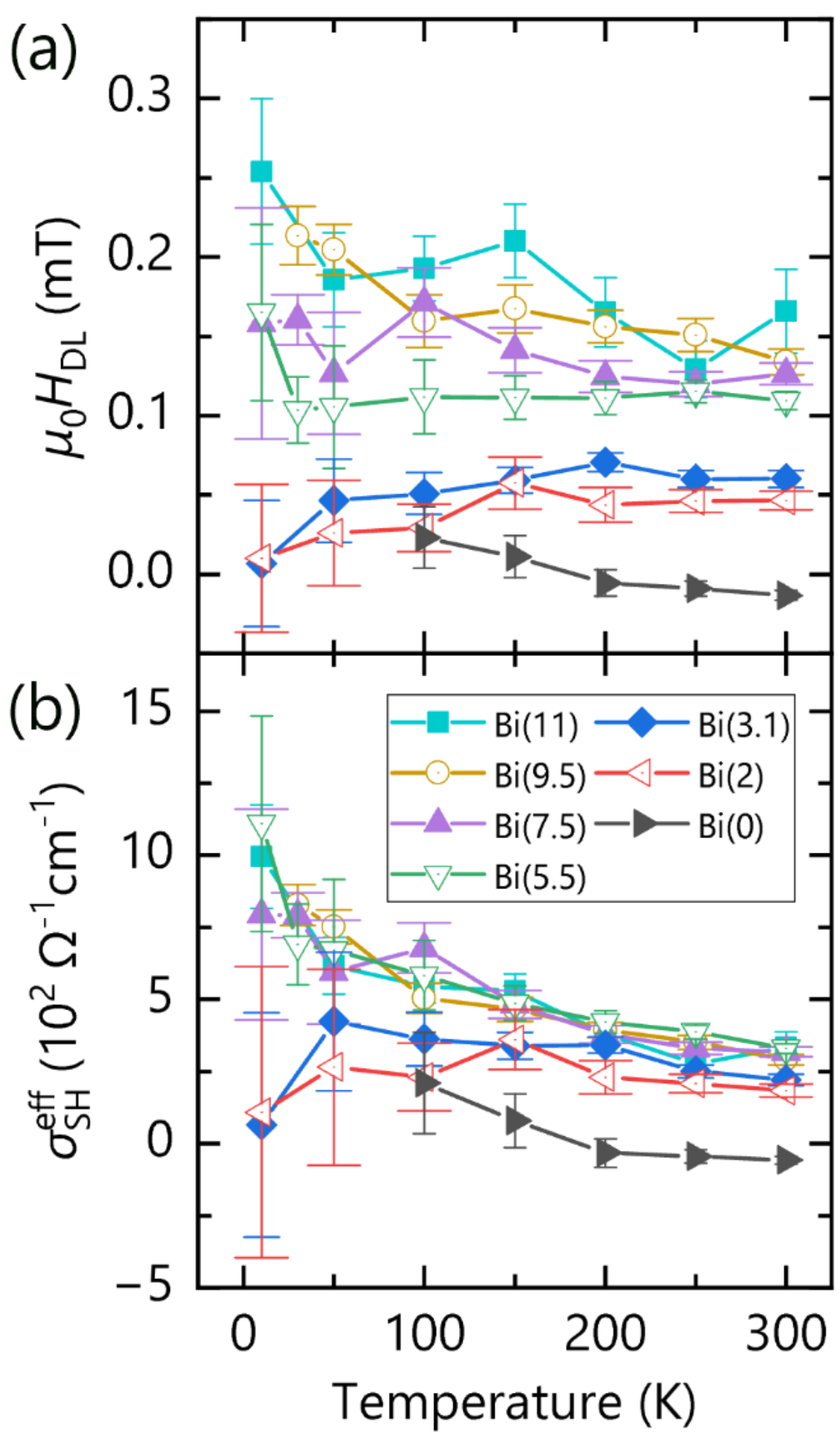

Fig. 4. Tatsuoka *et al*.

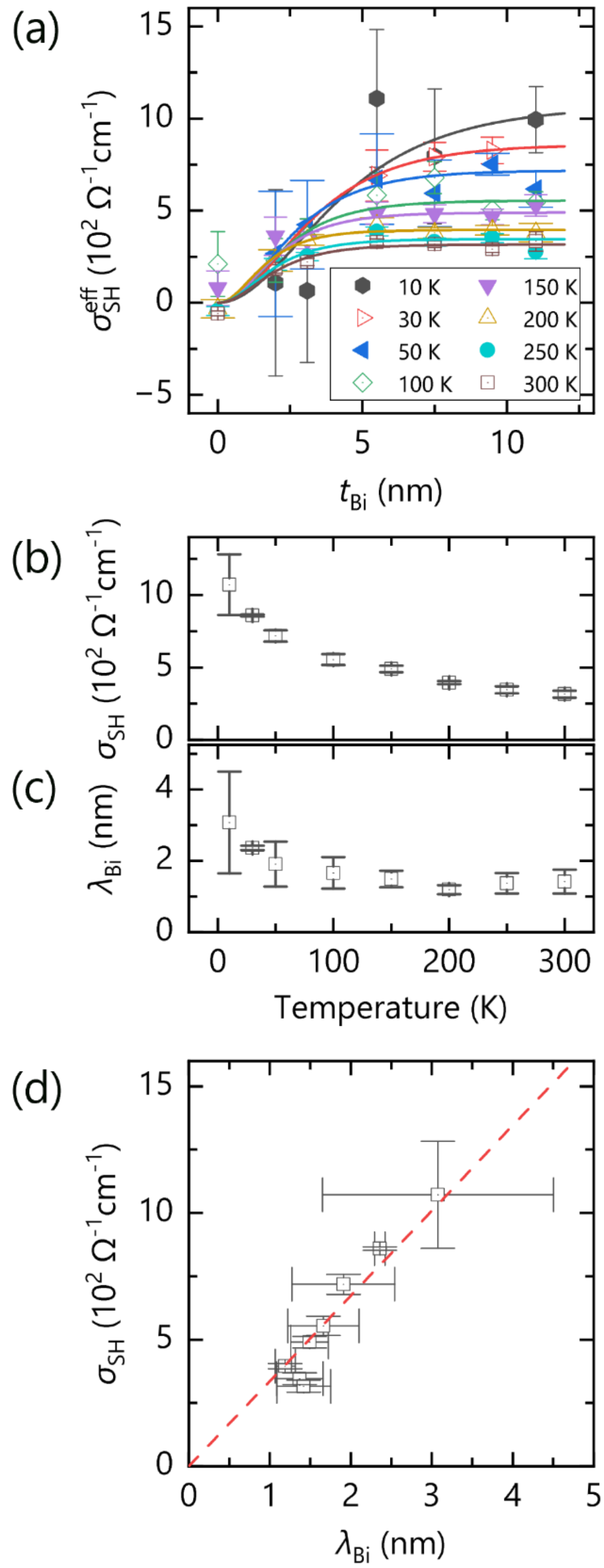

Fig. 5. Tatsuoka *et al*.